\documentclass[final,elsart1p,times,doublespacing]{elsarticle}
\usepackage{graphicx}
\usepackage{nicefrac}
\usepackage{amsmath}
\usepackage{amssymb}
\usepackage{textcomp}
\usepackage{color}
\newcommand{\red}[1]{  \textcolor{black}{#1}}

\biboptions{comma,sort&compress,numbers}
\journal{Carbon}
\begin{document}

\begin{frontmatter}

\title{Possible superconductivity in multi-layer-graphene by application of a gate voltage}
\author[exp]{A. Ballestar}
\author[exp]{P. Esquinazi\corref{corr1}}
\cortext[corr1]{Corresponding author. Tel/Fax: +49 341 9732751/69.
E-mail address: esquin@physik.uni-leipzig.de (P. Esquinazi)}
\author[exp]{J. Barzola-Quiquia}
\author[exp]{S. Dusari}
\author[exp]{F. Bern}
\address[exp]{Division of Superconductivity and Magnetism, Institut
f\"ur Experimentelle Physik II, Universit\"{a}t Leipzig,
Linn\'{e}stra{\ss}e 5, D-04103 Leipzig, Germany}
\author[cam]{R. R. da Silva}
\author[cam]{Y. Kopelevich}
\address[cam]{Instituto de F\'isica "Gleb Wataghin"/DFA, Rua Sergio
Buarque de Holanda Nr.777, Universidade Estadual de
Campinas-UNICAMP, Cidade Universit\'aria Zeferino Vaz, Bairro
Bar{\~a}o Geraldo 13083-859, Brasil}

\begin{abstract}

The carrier density in tens of nanometers thick graphite samples
(multi-layer-graphene, MLG) has been modified by applying a gate
voltage ($V_g$) perpendicular to the graphene planes. Surface
potential microscopy shows inhomogeneities in the carrier density
($n$) in the sample near surface region and under different values
of $V_g$ at room temperature. Transport measurements on different
MLG samples reveal that  under a large enough applied electric
field  these regions undergo a superconducting-like transition at $T
\lesssim 17$~K. A magnetic field applied parallel or normal to the
graphene layers suppresses the transition without changing appreciably the transition temperature.
\end{abstract}

%\pacs{}
\begin{keyword}
Multi-layer-graphene \sep electrical properties \sep superconductivity
\end{keyword}
\end{frontmatter}

\section{Introduction}
\label{Introduction}

Superconductivity in carbon based materials has been found in a
relatively large number of samples. Most of them are graphite
 based systems with a chemical doping. Intercalated
graphite compounds with doping elements as potassium
(C$_8$K)${\text{\cite{han65}}}$, Lithium
(C$_2$Li)${\text{\cite{bel89}}}$, Calcium
(C$_6$Ca)${\text{\cite{wel05}}}$ or Ytterbium
(C$_6$Yb)${\text{\cite{wel05}}}$ are found to be superconducting
with transition temperatures ($T_c$) from 1.9~K (for
C$_2$Li)${\text{\cite{bel89}}}$ up to 11.5~K (for
C$_6$Ca)${\text{\cite{wel05}}}$. Note that in all those cases the
superconducting state is obtained in materials with much less
anisotropy as pure graphite with its Bernal structure. Recently
published works indicate, however, the possibility of higher $T_c$
in graphite systems, namely: The magnetic response of
water-treated graphite powders suggests the existence of room
temperature superconductivity ${\text{\cite{sch12}}}$, supporting
the conclusion of old reports ${\text{\cite{ant74,ant75}}}$;
Measurements testing directly the embedded two dimensional
interfaces found within the graphite Bernal structure
 show evidence for the Josephson effect at temperatures clearly above 100~K$ {\text{\cite{bal13}}}$;
 Magnetization measurements performed on graphite samples with internal interfaces support the
 existence of a  superconducting-like behaviour \cite{schcar} at very high temperatures.

Previous work reported that sulfur-doped graphite composites shows
superconducting-like behavior up to 35~K
${\text{\cite{silvaprl}}}$. Recently published study
${\text{\cite{kaw13}}}$ reported that bringing alkanes into
contact with graphite surfaces triggers zero resistance at room
temperature. By doping graphite samples via phosphorous or argon
implantation, several superconducting-like steps in the resistance
vs. temperature were reported recently up to nearly room
temperature\footnote{Larkins, G., Vlasov, Y. Holland, K..
ArXiv:1307.0581}. We may therefore speculate that if the carrier
density increases in the graphene layers superconductivity might
be triggered. Although we should note that, taking into account
the clear difference between the critical temperatures obtained in
intercalated graphite compounds and those in doped  but not
intercalated  graphite, it appears that the critical temperature
increases the higher the anisotropy of doped graphite.

Most of the theoretical predictions about superconductivity in
graphite/graphene emphasize that it should be possible under the
premise of sufficiently high carrier density to reach $T_c > 1~$K
${\text{\cite{garbcs09,uch07,kop08}}}$. Following a BCS approach
in two dimensions energy gap values at 0~K of the order of $60~$K
have been obtained if the density of conduction electrons {\em per
graphene plane} increases to $n \sim 10^{14}~$cm$^{-2}$
${\text{\cite{garbcs09}}}$, in agreement with the theoretical
estimates based on different
approaches${\text{\cite{uch07,kop08}}}$. Also high temperature
superconductivity with a $d+id$ pairing symmetry has been
predicted to occur in doped graphene with a carrier concentration
$n \gtrsim 10^{14}~$cm$^{-2}$${\text{\cite{pat10}}}$. We note that
the intrinsic carrier density of defect-free graphene layers
inside graphite is $n \lesssim
10^9~$cm$^{-2}$${\text{\cite{arn09,dus11,neu09}}}$. However,
defects and/or hydrogen doping within regions at
interfaces${\text{\cite{bar08}}}$ or at the graphite surface may
show much larger carrier density, e.g. $n > 10^{11}~$cm$^{-2}$.
Therefore it is a challenge for experimentalists to increase the
carrier density above a certain threshold, at the interfaces or at
the regions that provide a coupling to those interfaces, to
trigger superconductivity. \red{The interfaces we are taking about
are quasi two-dimensional regions that are located between two
crystalline regions with Bernal stacking order each of them with a
slightly different angle respect to the $c-$axis
\cite{bar08,bal13}}.

We note that high temperature superconductivity has been also
predicted to occur at rhombohedral graphite surface regions due to
a topologically protected flat band ${\text{\cite{kop11,kop13}}}$
or in multilayered structures with hybrid stacking, i.e.
rhombohedral and Bernal stacking ${\text{\cite{mun13}}}$. In this
case, however, increasing the carrier density will not necessarily
increase $T_c$. A homogeneous doping strongly suppresses surface
superconductivity while non-homogeneous field-induced doping has a
much weaker effect on the superconducting order parameter
${\text{\cite{mun13}}}$. Therefore, the expected effect of an
electric field on the transport properties of inhomogeneous doped
multilayered graphite appears not so simple.

Recently published studies show that electrostatic  carrier
accumulation is an interesting tool to trigger new states of
matter at certain interfaces. In the particular case of graphite
Otani and collaborators predicted that nearly free electron states
distributed in quasi two dimensional ($2D$) regions at the
interfaces can cross the Fermi level if an external electric field
perpendicular to the graphene plane is applied
${\text{\cite{ota10}}}$. This opens the possibility of triggering
superconductivity in a pure carbon material. It is then appealing
to use the electrostatic charge doping to increase $n$ in graphite
without disturbing its quasi-$2D$ dimensionality. This expected
difference in critical temperatures  can be partially understood
within a BCS mean field model taking into account the role of
high-energy phonons in the $2D$ graphite structure
itself${\text{\cite{garbcs09}}}$. Metal decorated graphene samples
have been proven to have a tunable superconducting to insulating
transition via electric field gating
${\text{\cite{kes10},\cite{all12}}}$ where both chemical and
electrostatic carrier density doping are combined. However, no
experimental results for pure graphite  samples have been
published yet. Thus, the aim of the present study is to induce
large enough charge densities inside the MLG samples via electric
field gating without any chemical doping.

\section{Experimental details}
\subsection{Samples characteristics}
\label{samplecharacteristics}

The MLG samples were prepared from Highly Oriented Pyrolytic
Graphite (HOPG) flakes with the highest crystalline quality,
  ZYA grade (0.4$^\circ$ rocking curve width) from the former company
Advanced Ceramics. Small flakes from the as-received bulk piece of
HOPG were produced by peeling. These small flakes were used  to
produce the so-called multi-layer-graphene samples by a simple
rubbing procedure described elsewhere \cite{bar08} . This
procedure consists in a careful mechanical press and rubbing of
the initial material on a previously cleaned substrate. All the
samples were fixed on a 150 nm thick Si$_3$N$_4$ terminated
surface of a doped Si substrate. Samples with thicknesses between
20 to 90~nm (measured by Atomic Force Microscopy (AFM) and optical
microscopy) have been obtained. The electrical contacts were
prepared using electron beam lithography followed by Pt/Au thermal
evaporation,  {see Fig.1(a). The distance between voltage
electrodes varied between $\sim 1$ and 4$~\mu$m, upon sample.}
Samples quality has been checked by measuring Raman spectra (see
Supporting information).

 Regarding the possible influence of the used preparation
procedure on the existence and distribution of the superconducting
regions in the studied samples, we note the following. In
\cite{bar08} it has been shown that the temperature dependence of
the resistance of samples obtained from the same HOPG bulk sample
as used in this work, is related to the
existence of internal interfaces. Neither Raman nor the transport measurements
of several samples prepared with the rubbing method
indicate any relevant influence.  The interfaces,
which existence has been known for a relatively long time
\cite{ina00}, are found in some, not all, HOPG samples.
The superconducting properties due to
these interfaces remain, independently of the method used to
prepare the samples obtained from the same bulk HOPG
\cite{bal13,sru11}. Therefore, there is no clear experimental
evidence  that speaks for or against an influence of the rubbing
method on the transport characteristics of the samples.

\begin{figure*}[]
%\vspace{0.5cm}
\begin{center}
\includegraphics[width=1.\columnwidth, angle=0]{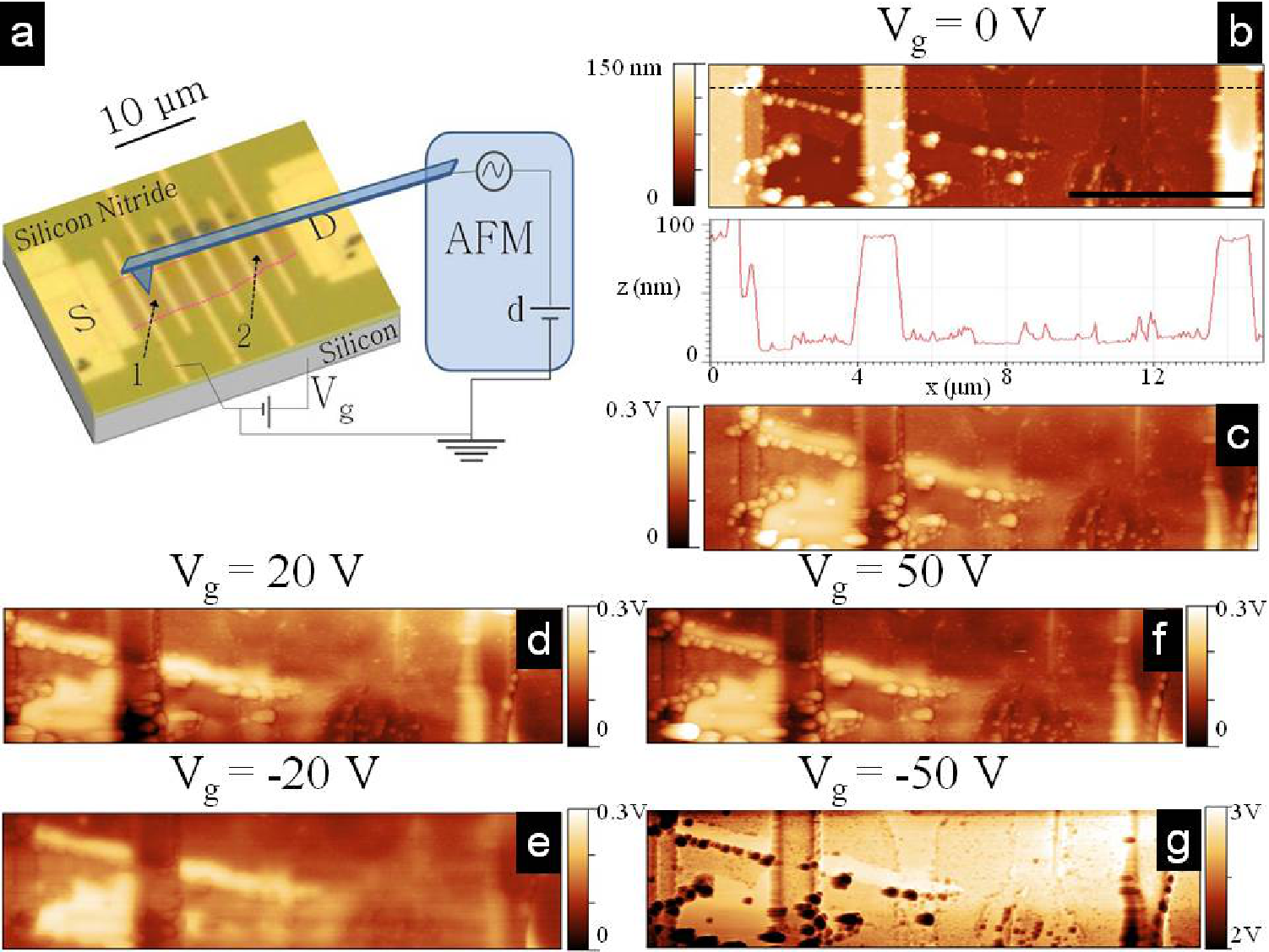}
\caption[]{(a) Sketch of the configuration used for the surface
potential measurements, where the AFM setup is presented,
including the recorder of the voltage variation coming from the
studied sample (indicated with ``d" in the small blue color
circuit sketch). A distance of 50~nm between AFM tip and sample
surface has been used in all the presented  images. The
configuration used in all the measurements shown in this work with
the gate voltage supply $V_g$ is also shown. The optical
microscope picture shows an image of sample S5, where drain (D)
and source (S) contacts are indicated as well as two studied areas
($1$ and $2$) (between two electrical contacts) selected to obtain
the surface potential differences. (b) AFM topography image of
sample S5. The black scale bar at the bottom right represents
5~$\mu$m length. The picture below shows the profile scanned along
the dashed line region in (b). (c) to (g): Images showing the
measured surface potential obtained at the indicated $V_g$ values
at room temperature in the area~1. The picture (c) corresponds to
0~V.}\label{AFM}
\end{center}
\end{figure*}

The samples surfaces have been studied by Atomic Force Microscopy
(AFM). We used a Dimension 3000 scanning probe microscope with a
Nanoscope IIIa controller and phase extender (Digital Instruments
Inc., Santa Barbara, CA). A sketch of the experimental setup is
shown in Figure~\ref{AFM}(a) where the AFM and the gate voltage
supply are presented. Note that both devices have common ground
and that the configuration used allows us to create an external
electric field perpendicular to the graphene planes. The gate
voltage was applied between the conducting doped Silicon substrate
and the sample surface through the large electrical resistance of
the 150~nm Si$_3$N$_4$ layer. Every experiment has been operated
within the non-breaking range of this insulating layer.

An optical microscope image of sample S5 is shown in
figure~\ref{AFM}(a); numbers $1$ and $2$ refer to different
studied areas. In figure~\ref{AFM}(b) an AFM image of the
topography of a part of this sample ($15~\mu$m~$\times 3.75~\mu$m)
is presented. The brightest areas correspond to the electrical
contacts and the two darker stripes in the middle of the image
correspond to two destroyed electrical contacts. Besides that, the
sample appears to be flat.  {As we are interested in studying the
carrier density distribution in the near surface region we have
performed surface potential measurements, also known as Kelvin
probe microscopy (KPM)${\text{\cite{non91}}}$}. In this operating
mode we record the voltage on the sample surface in the following
way. In a first scan the sample topography is recorded. In a
second scan at a height of 50~nm above the sample surface the
Coulomb interaction between tip and sample is eliminated by
applying a voltage equal to the difference of the work-functions
of sample and tip controlled by a feedback loop.  In principle, we
may expect that voltage variations on the sample indicate
different Fermi levels and thus different carrier densities $n$ as
the whole sample is a piece of multigraphene, apart from the gold
contacts.

Figure~\ref{AFM}(c) shows the phase image obtained without any
applied gate voltage ($V_g$ = $0$V).  One can easily realize that
the obtained signal from the sample surface is not homogeneous,
i.e. it shows a location dependent work-function. It can clearly
be seen also that different potentials are closely related to
different surface features. The most probable scenario is the one
related to the surface filth or adsorption of molecules proposed
and measured in \cite{mar13} . Figures~\ref{AFM}(d) to (g) show
the effect of different $V_g$ on the sample surface potential. For
$V_g \pm 20~$V (Figures~\ref{AFM}(d) and (e)), no remarkable
change with respect to no applied electric field is observed. No
essential effect is observed at higher positive $V_g$ (see
Figure~\ref{AFM}(f) obtained at $V_g = 50~$V). However, if we use
$V_g = -50~$V, see Figure~\ref{AFM}(g), we recognize a clear
change where brighter areas are detected with a larger potential
variation within the measured region. Taking into account the
scales on the right of each image, we can see that in the case of
Figure~\ref{AFM}(g) the variation is $\gtrsim 0.5~$V, while in the
rest of the cases is $\lesssim$~0.3~V.

 The important message for this work   is that different
sample regions apparently react differently, i.e. the electrical
field induces heterogeneous doping pointing to intrinsic
variations of the electronic structure. This provides us a way to
understand the non percolative superconducting transition we
describe below. It can be understood taking into account that an
ideal graphite matrix with Bernal stacking is
semiconducting${\text{\cite{gar12}}}$ and that internal
interfaces${\text{\cite{bar08,bal13,schcar}}}$ plus other
defects${\text{\cite{arn09}}}$ affect the effective carrier
density and therefore the screening characteristics in specific
regions. Summarizing, we observed an inhomogeneous surface
potential response, which reacts asymmetrically to the electric
fields applied perpendicular to the graphene planes.

In order to further investigate the effect of the gate voltage on
the carrier density of MLG samples we studied the transport
properties, i.e. resistance behavior as function of temperature
and magnetic field. We use a DC input current of $1~\mu$A supplied
by a Keithley 6221 current source and the voltage was measured
with a Keithley 2182 DC nanovoltmeter,  {always using the usual
four probe method}. When we apply a gate voltage the configuration
shown in Figure~\ref{AFM}(a) is the one used for all the samples.

\red{The measurements as a function of temperature and magnetic
field were performed in a He-flow cryostat (Oxford Instruments) in
the temperature range between 2~K and 250~K with temperature
stabilization better than 1~mK. The magnetic field was applied
with a superconducting solenoid in permanent modus. Its value was
obtained from a previously calibrated Hall sensor located on the
sample holder. The magnetic field was applied perpendicular as
well as parallel to the main area of the samples (i.e. graphene
planes) using a step-motor controlled sampler holder rotation
system. The measurements were done as follows: first, we applied
the corresponding $V_g$ at 2~K and after a stable value of the
resistance was reached a certain magnetic field was applied. Then,
the resistance $R(T)$ was measured from 2~K to 25~K. Once the
$R(T)$ curve was measured, the magnetic field was set to zero and
the process was repeated from 2~K with other set field. The gate
voltage has been maintained during the whole set of measurements.}

\begin{figure}[]
%\vspace{0.5cm}
\begin{center}
\includegraphics[width=1.05\columnwidth]{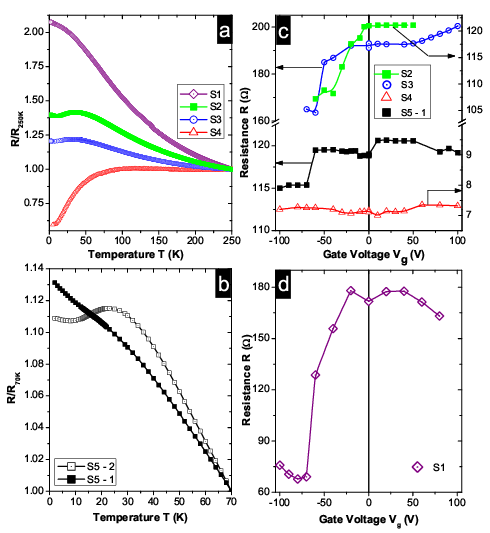}
\caption[]{(a) Temperature dependence of the normalized resistance
at zero gate voltage and zero magnetic field of four samples.
\red{The sample S4 had the largest thickness (90~nm), whereas the
other samples had a thickness 30-45, 22, and 30-45~nm for S1, S2
and S3, respectively}. (b) The same but for the two different
regions $1$ and $2$ of sample S5 (thickness 20-35~nm), see
Figure~\ref{AFM}(a). (c) Gate voltage dependence of the resistance
at 2~K for samples S3 and S5-1 (left $y-$axis) and S2 and S4
(right $y-$axis). (d) Gate voltage dependence of the resistance of
sample S1 at 2~K.} \label{T-dep}
\end{center}
\end{figure}

The temperature dependence of the resistance $R(T)$ at zero gate
voltage is shown in Figures~\ref{T-dep}(a) and (b). We present
here the results of five samples named S1, S2, S3, S4 and S5,
showing slightly different behaviors. S1, S2 and S3 samples
(30-45, 22, 30-45~nm thickness) show a semiconducting behavior
intrinsic to the graphite Bernal structure influenced partially by
lattice defects, especially $2D$
interfaces${\text{\cite{gar12}}}$. Sample S4 (90~nm thickness)
shows a clear metallic behavior below 100~K ascribed to its higher
number of internal interfaces ${\text{\cite{bar08}}}$. The level
off of $R(T)$ of sample S1 below 25~K and the features below 50~K
in samples S2 and S3, are also related to the role of these
interfaces and/or to the contributions of the free surface of the
sample (or sample-substrate interface), as discussed in Refs.
\cite{bar08} and \cite{gar12}. The results for sample S5
(non-homogeneous thickness 20-35~nm) obtained at the two different
regions labeled 1 and 2 in Figure~\ref{AFM}(a) are shown in
Figure~\ref{T-dep}(b). While in the region $1$ the semiconducting
behavior remains even at low temperatures, the behavior of the
sample area $2$ reveals a similar level off or maximum as in
samples S1, S2 and S3, which is related to the contribution of the
internal interfaces and other lattices defects in the sample
regions where the voltage electrodes are located. The clear
difference in the $R(T)$ function within a few microns distance
within the same sample is in qualitative agreement with the
surface potential microscopy results shown previously as well as
similar transport results obtained in thin graphite
samples${\text{\cite{arn09}}}$ and earlier EFM results on bulk
HOPG sample surfaces${\text{\cite{lu06}}}$ that revealed
sub-micrometer domain like carrier density distributions in
graphite surfaces.

\subsection{Electrostatic screening in MLG samples}
\label{screening}

Different theoretical approaches calculated the distribution of
charge under an external electrical field perpendicular to the
graphene planes in graphite${\text{\cite{vis71},\cite{gui07}}}$.
Using a random phase approximation, depending on whether the
inter-layer electron tunneling was taken into account or not,
screening lengths between $\lambda=0.54~$nm${\text{\cite{vis71}}}$
and $\lambda=0.7$~nm${\text{\cite{gui07}}}$ have been obtained.
However, Miyazaki and collaborators experimentally measured a
screening length of $1.2 \pm 0.2$~nm ${\text{\cite{miy08}}}$,
which corresponds to 3 or 4~graphene layers. Kuroda and coworkers
theoretically found that the actual screening depth depends on the
experimental conditions, in particular the actual doping of the
sample and the temperature${\text{\cite{kur11}}}$. Furthermore,
they found that a variation of more than an order of magnitude can
be obtained. We use their model to estimate the penetration depth
of the applied electric field inside our samples.

The intrinsic carrier density in each graphene layer in the
graphite structure, without defects and interfaces, should be $n <
10^{8}~$cm$^{-2}$ at the temperature of our experiments
${\text{\cite{arn09},\cite{dus11},\cite{esq11}}}$. The exact value
of $n$ for the graphene layers in each of the measured samples is
not really well known just because the samples are not free from
defects and impurities as, e.g., hydrogen. We take as upper limit
$n \sim 10^{8}~$cm$^{-2}$ for the graphene layers not involved in
the internal interfaces or at the surface. Following \cite{kur11}
we estimate that at $T < 30~$K the effective penetration depth
should be equal to at least 7~graphene layers or about 2.4~nm.
That would mean that the electric field mainly influences the near
surface region of the samples, in case that no internal interfaces
with much larger carrier density exist. Otherwise, if the carrier
density is larger, the screening effect will be more relevant. Our
estimate is basically in agreement with other theoretical work,
which showed that the electric field should be screened within a
few layers from the sample surfaces ${\text{\cite{kos10}}}$. In
case the carrier density of the non-defective graphene layers is
smaller, the larger will be the penetration depth of the electric
field in the sample.

\section{Results}

\subsection{Gate voltage effect}
\label{gatevoltage}

Although few layer graphene systems have been extensively studied,
the properties of graphene-based systems with a higher number of
layers ($\sim 10$ or more) are still a matter of discussion.
Particularly, the gate voltage effect is not fully clarified
probably because the complexity related to the screening effect,
the dependence on the number of layers and the corresponding
presence (or absence) of interfaces with higher carrier
densities${\text{\cite{bar08},\cite{gar12},\cite{bal13}}}$ as
well as further inhomogeneities due to lattice defects and
impurities. These facts make essentially every sample slightly
different from the others and even different regions within the
same sample can show noticeable differences concerning the
electronic properties, as the results for sample S5 in
Figure~\ref{T-dep}(b) demonstrate (see also \cite{arn09}
particularly the results in Figure~4).

Due to the electrostatic screening in MLG samples, it is essential to
apply a large amplitude of the electric field on the sample in
order to see an effect on the electrical resistance. Typical gate
voltage values for MLG samples on SiO$_2$ of 300~nm thickness
substrates are 100~V${\text{\cite{nag10}}}$. Otani and coworkers
calculated an electric field of 0.49~V/$\AA$ in order to inject
free-electron carriers at the Fermi level${\text{\cite{ota10}}}$. As the
number of layers in our samples is larger than in their work, we
use back gate voltages up to 100~V.

Figures~\ref{T-dep}(c) and (d) show the resistance vs. applied
gate voltage $V_g$ at a constant temperature of 2~K for samples S1
to S5, this last in  region $1$. With  exception of the
thickest sample S4, all samples show  an asymmetric behavior
respect to zero voltage with a clear decrease of the resistance at
certain negative $V_g$'s. For positive voltages the resistance
either increases slightly or it does not change significantly. We
think that these differences, as well as that in the relative
decrease of resistance at a given negative gate voltage, are
related to the overall inhomogeneities of the MLG samples due
to impurities or to the higher carrier densities located at the internal
interfaces of the samples. The absence of any significant change
with $V_g$ in the thickest sample S4 is a clear indication for the
screening effect of the electric field. A similar behavior of the
resistance of MLG samples with the applied gate voltage was
partially reported by Kim and collaborators${\text{\cite{kim05}}}$.
However, in that work no results on the temperature or the
magnetic field dependence of the resistance under a gate voltage
were reported.

\begin{figure}[]
\begin{center}
\includegraphics[width=1\columnwidth]{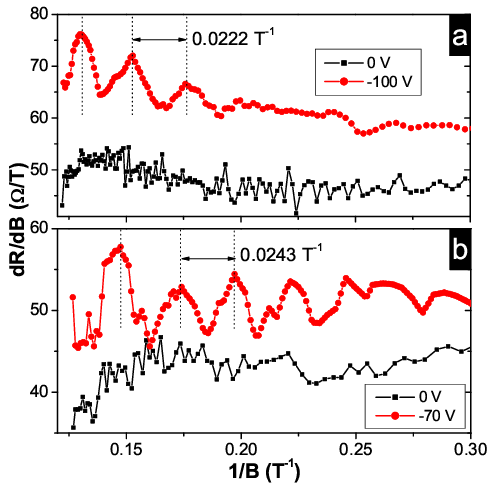}
\caption[]{First derivative of the magnetoresistance vs. inverse
field at 2~K and with (red open circles) and without (black solid
squares) applied gate voltage of (a) sample S1 and (b) sample S5
region $2$.}\label{SdHNew}
\end{center}
\end{figure}

The clear drop observed in the resistance for large enough
negative $V_g$ in different MLG samples suggests the existence of
a superconducting-like transition that should be also recognized
as a function of temperature, as shown below. The overall change
of the resistance with $V_g$ indicates that there might be an
increase of the carrier density at the Fermi level in some parts
of the sample. A way to check whether there is a real increase in
the carrier density with applied gate voltage is to measure the
Shubnikov-de Haas (SdH) oscillations of the magnetoresistance.
Figure~\ref{SdHNew} shows the first field derivative of the
magnetoresistance vs. inverse field at 2~K with and without
applied gate voltage corresponding to sample S1 (a) and  sample S5
region $2$ (b), as examples. At no applied gate voltage the data
reveal no SdH oscillations in the shown field range and within
experimental error. This is actually expected because ideal
graphite is a narrow band semiconductor and at low temperatures no
Fermi surface should exist${\text{\cite{gar12}}}$. At a large
enough applied gate voltage clear SdH oscillations are observed in
the first field derivative for fields above $\sim 4~$T. From the
obtained period of the oscillations we estimate a $2D$ carrier
density $n \simeq 2.2 \times 10^{12}~$cm$^{-2}$ for S1 and $n
\simeq 2 \times 10^{12}~$cm$^{-2}$ for S5 at region $2$, one order
of magnitude larger that the one obtained for the bulk graphite
sample${\text{\cite{arn09}}}$ from which the MLG samples were
obtained.

We stress that the carrier density obtained for the bulk sample is
not intrinsic of the Bernal graphite structure but it is related
mainly to the internal interfaces (or other defective
regions${\text{\cite{arn09},\cite{gar12}}}$) commonly found in the used highly
oriented pyrolytic graphite samples${\text{\cite{bar08}}}$. We note also
that  the rather weak SdH oscillations are observed
only at fields above 1~T indicating the existence of domains of
size $2r_c < 100~$nm in which $\lambda_F < 50~$nm, i.e. domains
with $n
> 10^{11}~$cm$^{-2}$ within a matrix of much lower carrier
concentration (i.e. $n \simeq
10^{-9}~$cm$^{-2}$${\text{\cite{neu09}}}$) ($r_c$ and $\lambda_F$
are the cyclotron radius and Fermi wavelength,
respectively)${\text{\cite{arn09}}}$. Therefore, we can assume
that, if the distance between the high carrier density domains is
about the same as the domains size, the applied gate voltage
couples these high carrier density localized domains. In this case
the decrease of the resistance observed in Figures~\ref{T-dep}(c)
and (d) for large enough negative $V_g$ can be understood.

\begin{figure}[]
%\vspace{0.5cm}
\begin{center}
\includegraphics[width=1\columnwidth]{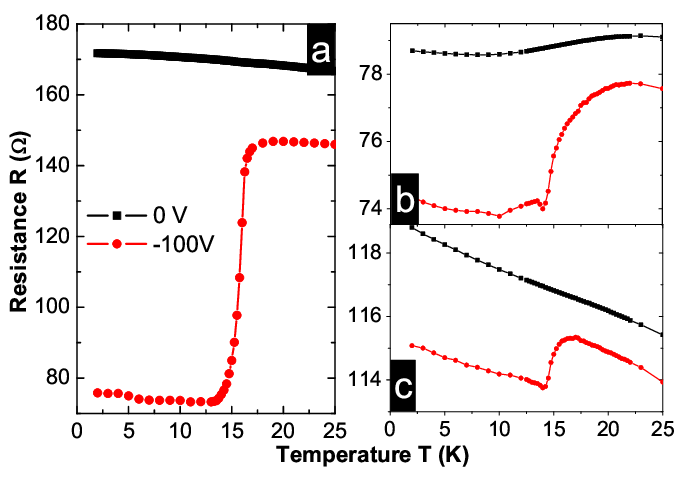}
\caption[]{(a) Temperature dependence of the resistance of samples S1
 and (b) S5 for region $2$ and (c) region $1$ at no gate applied
(solid black squares) and $V_g = - 100~$V (solid red
circles).}\label{RvsTVgon}
\end{center}
\end{figure}

In order to further investigate the effect of the gate voltage on
the electronic properties, the temperature dependence of the
resistance under a constant gate voltage has been studied.
Figure~\ref{RvsTVgon} shows the results obtained for samples S1
and S5 without applied gate  voltage and with $V_g = -100$~V. A clear step
like transition below 20~K appears if $V_g$ is applied. The
details of the transition depend on the selected sample (see
Supporting information). Even the two studied regions in sample S5
present some differences (see Figure~\ref{RvsTVgon} (b) and (c)),
again a sign of the existence of inhomogeneities in the samples.
The observed transition at $T\lesssim20~$K indicates that after
the application of a negative large voltage the large carrier
density $n$ located in some parts of the samples (as shown in the
SdH oscillations in Figure~\ref{SdHNew}) induces either
superconductivity in those parts or they provide a kind of
Josephson link between superconducting regions already existent in
the samples at certain interfaces${\text{\cite{bal13}}}$. Which of these
two possibilities is  the correct description can be
answered measuring the magnetic field dependence of the
resistance, as we show below. For both cases it applies that  the
reason for non percolation, i.e. non zero resistance in the presumable
superconducting state, is simply related to the fact that the
voltage electrodes are not contacting the superconducting regions
directly.

\begin{figure}[]
%\vspace{0.5cm}
\begin{center}
\includegraphics[width=1.0\columnwidth]{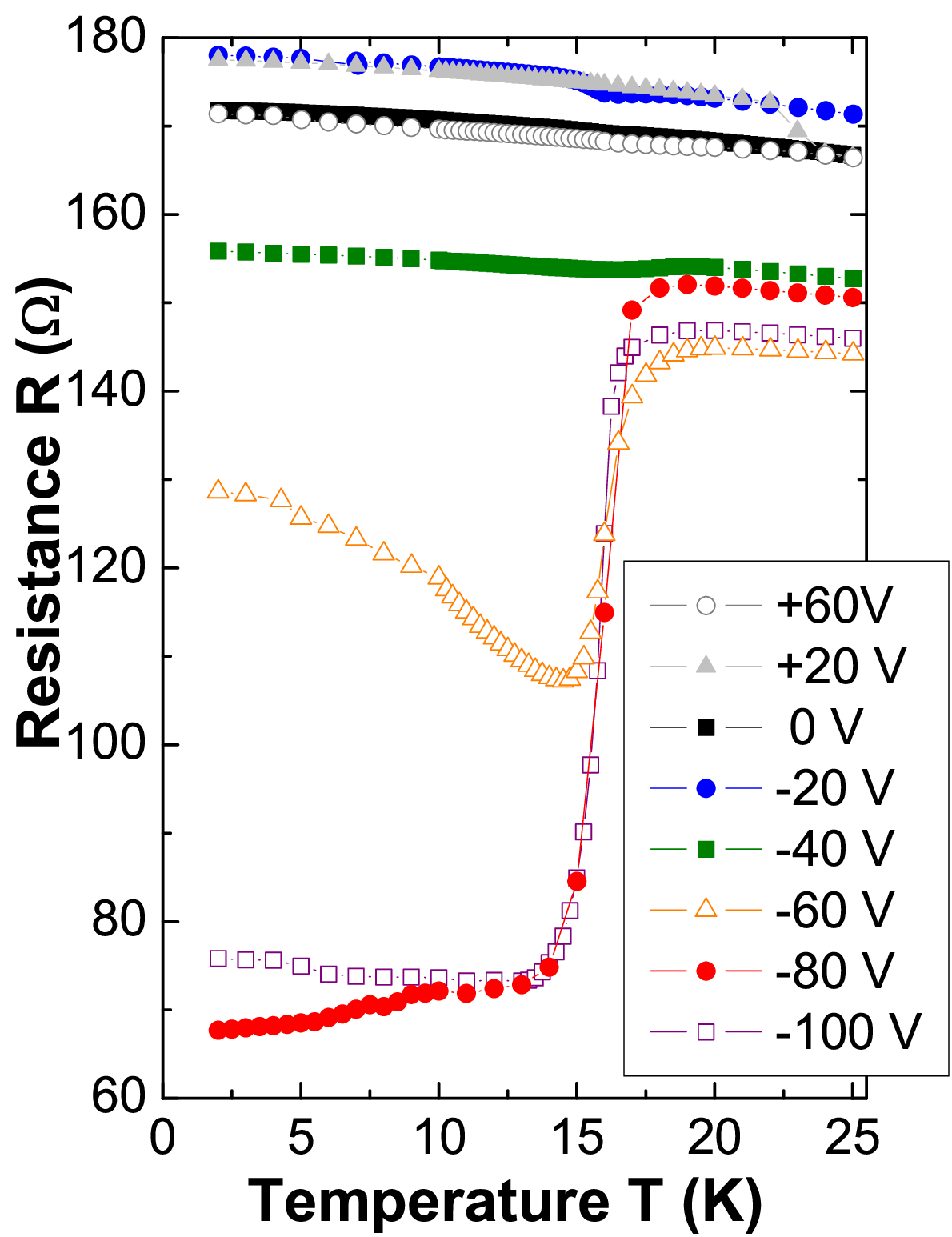}
\caption[]{Temperature dependence of the resistance of sample S1
at different constant values of $V_g$.}\label{Fig5}
\end{center}
\end{figure}

The effect of different gate voltages on the temperature
dependence of the resistance was also investigated (see
Figure~\ref{Fig5}) and the asymmetric behavior of the resistance
with $V_g$ (see Figure~\ref{T-dep}) is recognized. The temperature
dependence of the resistance does not change significantly for
$-40V < V_g < +60$~V. \red{The change in resistance produced at
-20~V~$\le V_g \le 60~$V remains small (less than 5\%) and it is
not monotonous, see Fig.~\ref{Fig5}}. At $V_g=-40$~V a small dip
appears at $T \simeq 17$~K. For $V_g < -40$~V the small dip
develops in a clear step with less than 2~K transition width. The
curve obtained at $V_g=-60$~V shows a drop of the resistance at
15~K and an upturn below it. This indicates the existence of a
non-uniform channel with superconducting regions connected in
series with normal ones. As shown in Figure~\ref{Fig5} for sample
S1, the lower the gate voltage the clearer the transition. Note
that the transition temperature does not change with the applied
gate. This general behavior observed in all samples suggests that
the increase in carrier density is not really triggering a
superconducting transition but it enables a link between the
already existent superconducting regions. This result would
indicate that field-induced superconductivity should not be
observed in single graphene or MLG samples without interfaces (or
the regions where superconductivity is localized).

\subsection{Magnetic field effect under a finite gate voltage}
\label{magneticfield}

\begin{figure}[]
%\vspace{0.5cm}
\begin{center}
\includegraphics[width=0.95\columnwidth]{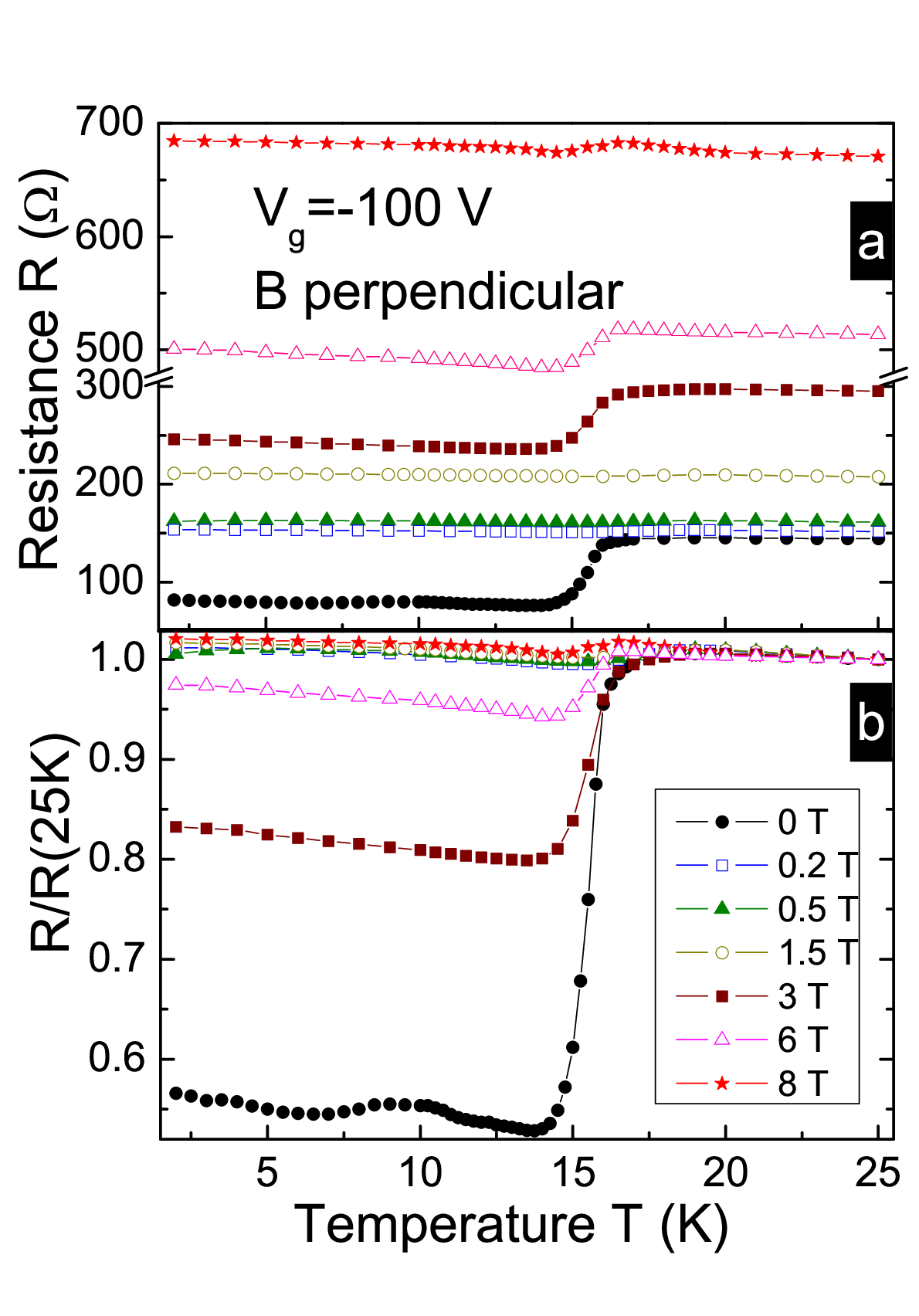}
\caption[]{Temperature dependence of the resistance of sample S1
at different constant values of magnetic field applied perpendicular to
the graphene planes and at $V_g = -100~$V. (a) Absolute values of
the resistance. (b) Normalized values of the resistance by its value
 at 25~K vs. temperature at the same applied fields as
in (a).} \label{RvsTBn}
\end{center}
\end{figure}

For a better characterization of the nature of the observed
transition under an applied electric field we need to study the
magnetic field effect on it. We have measured therefore the
temperature dependence of the resistance under a large enough gate
voltage at constant magnetic fields. In what follows we discuss
mainly the results of sample S1 for both, magnetic field applied
normal and parallel to the graphene planes. Results for the other
samples are qualitatively similar. Figure~\ref{RvsTBn} shows the
dependence of the resistance  with temperature at different
applied fields normal to the main area of the sample and at a
fixed gate voltage of -100~V. As expected for this field
direction, the background resistance increases with field due to
the usual magnetoresistance of the MLG samples, see
Figure~\ref{RvsTBn}(a). \red{To suppress the effect of the
magnetoresistance contribution and show clearer the effect of the
magnetic field on the transition, Figure~\ref{RvsTBn}(b) shows the
same data as in (a) but normalized.}. As shown in
Figure~\ref{RvsTBn} a field of 0.2~T is enough to suppress
completely the transition at 17~K. This suppression remains to a
field of 1.5~T. At a field of 3~T and higher the transition
appears again at the same temperature but slightly broader and it
nearly vanishes at 8~T, see Figure~\ref{RvsTBn}(b). Note that the
temperature of the transition does not change significantly with
applied field. The transition is rather unconventional because the
magnetic field affects mainly the relative step height of the
transition. This fact also suggests that the field does not affect
the superconducting regions themselves but mainly the coupling
between them produced by the applied electric field,
\red{indicating also that the upper critical field would be higher
than 8~T. A similar conclusion can be taken from the increasing
difference between FC and ZFC magnetic moment data at high-fields,
see Fig.~6 in the supporting information of Ref.~\cite{sch12}.}

\begin{figure}[]
%\vspace{0.5cm}
\begin{center}
\includegraphics[width=1\columnwidth]{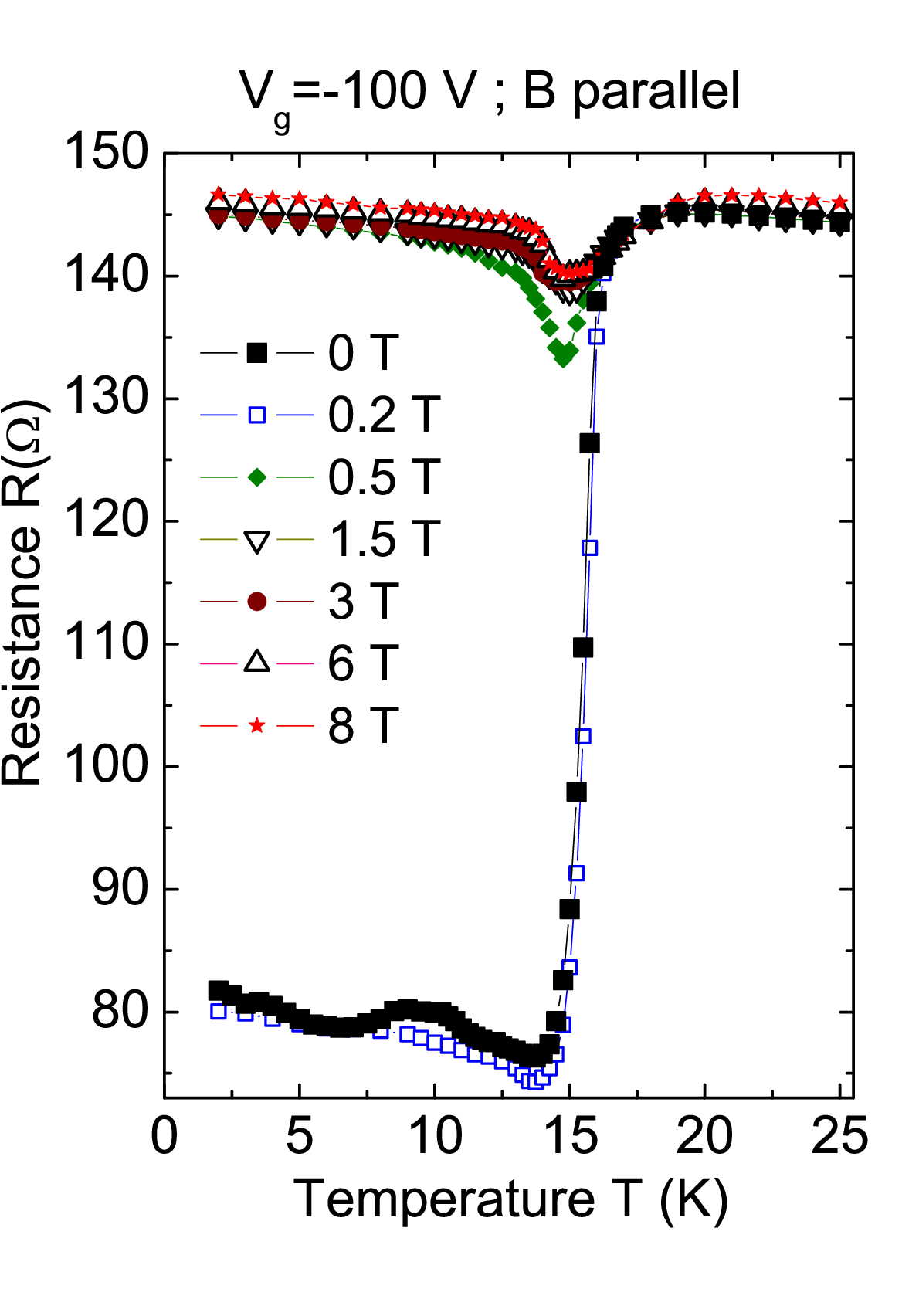}
\caption[]{Temperature dependence of the resistance of sample S1
at different constant values of the field applied parallel to the
graphene planes and at $V_g = -100$~V.} \label{RvsTBp}
\end{center}
\end{figure}

Figure~\ref{RvsTBp} shows the resistance vs. temperature at
 $V_g = -100$~V and at different magnetic fields applied
parallel to the graphene planes of the sample. The misalignment of
the field is less than $0.5^\circ$. The measured data show that
the transition remains unaffected by a parallel field of 0.2~T, in
contrast to the normal field result. At higher fields, however, it
is suppressed monotonously without any sign for a reentrance. Note
that the resistance above $\simeq 17$~K does not change
practically with field in agreement with the fact that the
magnetoresistance of graphite depends mainly on the normal field
component to the graphene planes${\text{\cite{kempa03}}}$. This
indicates clearly the absence of any Lorenz-force driven effect or
a change in the electron system at all above the critical
temperature. As for the normal applied fields, for parallel
applied fields the transition does not shift significantly to
lower temperatures and the resistance shows a minimum just below
the transition.

\begin{figure}[]
%\vspace{0.5cm}
\begin{center}
\includegraphics[width=1\columnwidth]{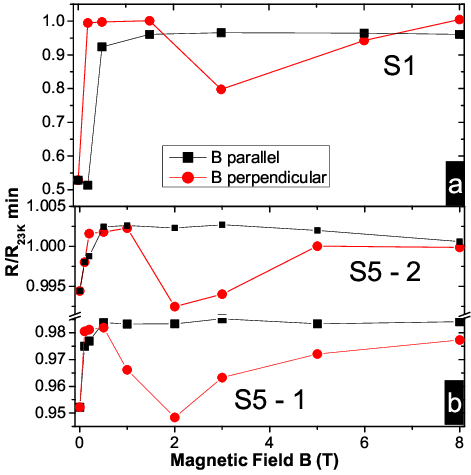}
\caption[]{Change of the resistance minimum normalized by the
resistance value at 23~K for sample S1 (a) and S5 (areas $1$ and
$2$) (b) with the magnetic field applied normal (empty blue
circles) and parallel (solid red squares) to the graphene planes
at $V_g = -100$~V.}\label{RminvsB}
\end{center}
\end{figure}

The magnetic field behavior of the transition for both field
directions, i.e. the resistance below the transition increases
with field without a clear decrease of the transition temperature
within the used field range, suggests that a filamentary
superconducting path  produced by the applied gate voltage is
affected by the magnetic field. In Figure~\ref{RminvsB} we compare
the results for both field directions by plotting the minimum
resistance just below the transition normalized by the resistance
value at 23~K vs. applied field for samples S1 and S5 in both
studied regions. Although some differences appear between S1 and
S5, we obtain qualitatively the same results, i.e., a small
perpendicular field of 0.1T is enough to vanish the transition and
a reentrance is observed at high enough applied magnetic field
normal to the graphene layers. The rather weak anisotropy of the
low field necessary to affect the transition peaks for the
triggering of a $3D$ filamentary path by the electric field.

The main difference in the behaviors obtained as a function of
field direction is related to the reentrance observed only for
fields normal to the graphene planes. The reentrance of the
transition observed for this configuration appears to have an
orbital character. We note that a similar effect has been observed
through the measurement of the conductance of a high-mobility 2D
electron gas between superconducting contacts at high fields
applied normal to the main 2D area${\text{\cite{moo99}}}$. This
reentrance or increase in the conductance with magnetic field was
explained arguing the increase in the probability of Andreev
reflections above a certain field${\text{\cite{moo99}}}$. A comparison
of our results with those from  \cite{moo99} is
permissible because the carriers mobility in the graphene layers
of our samples is huge${\text{\cite{dus11}}}$ and indications for the
influence of Andreev reflections in the magnetoresistance have
been also reported in similar MLG${\text{\cite{esq08}}}$.

\section{Conclusion}

We studied the behavior of the resistance of several MLG samples
as a function of temperature and magnetic field and under the
influence of a gate voltage applied normal to the graphene planes.
Taking into account relevant literature on granular
superconductors as well as the one obtained recently for the
internal interfaces in graphite samples, it appears natural to
assume that the transition in the resistance that develops at $T
\sim 17$~K with negative gate voltage is related to a
non-percolative superconducting-like state. Several open questions
remain, as for example why the apparent transition temperature is
about 17~K and not at much higher temperatures, as measured from
direct measurements of the response of the embedded interfaces in
graphite lamellae${\text{\cite{bal13},\cite{schcar}}}$ or in water
treated graphite powders ${\text{\cite{sch12}}}$.  A possible
answer to this important question is probably related to the
electric field triggering of the $3D$ (and not $2D$) connecting
paths between the already superconducting regions, these last much
less influenced by the electric field. In this case the $3D$
superconductivity temperature in graphite should be much near the
$3D$ graphite intercalated
compounds${\text{\cite{han65},\cite{wel05}}}$ than in the
discovered $2D$ superconductivity. The observed magnetic field
effects would influence the connecting paths and not the intrinsic
superconducting regions. Three dimensional paths and not only $2D$
appear necessary in order to explain the weak field anisotropy.
Finally, we would like to note the report on gate-induced
superconductivity in carbon nanotubes${\text{\cite{yan12}}}$ at
temperatures above 12~K, results that support the findings of
these studies.

\section*{Acknowledgement}
This work was supported by the Deutsche Forschungsgemeinschaft
under contract DFG ES 86/16-1 and by the DAAD project Nr.56269524
under PROBRAL and CAPES. R.R.d.S. and Y.K. acknowledge the support
from FAPESP, CNPq, CAPES-PNPD 1571/2008, ROBOCON, and INCT
NAMITEC. A.B. and S.D. were supported by the Graduate School of
Natural Sciences "BuildMoNa" of the University of Leipzig and
ESF-NFG and ESF ``Energie" from the European Fonds for the state
of Saxony.

%\bibliography{C:/Users/Ana/Desktop/AdvFuncMat/magneticcarbon}

%\bibliography{/Users/pablo/Dropbox/Documents/magnetic_carbon}

%\bibliography{/Users/pablo/Dropbox/Documents/Bias Voltage/magneticcarbon}
%\bibliography{D:/DATA/Dropbox/Documents/magnetic_carbon}
%\bibliography{/Users/pablo/Documents/Bias voltage/Adv Mate Version/new version/magneticcarbon}
%\bibliography{E:/Dropbox/Documents/Bias Voltage/magneticcarbon}
\bibliographystyle{model3-num-names}

%\bibliography{/Users/pablo/Dropbox/Documents/magnetic_carbon}
%\bibliography{E:/Dropbox/Documents/magnetic_carbon}

\end{document}